\pdfoutput=1
\documentclass[11pt]{article}

\usepackage[utf8]{inputenc}
\usepackage[T1]{fontenc}
\usepackage[margin=1in]{geometry}
\usepackage{graphicx,amsmath,amssymb,bm,placeins}
\usepackage[numbers,sort&compress]{natbib}
\usepackage{xcolor}
\usepackage{authblk}
\usepackage{caption}
\usepackage[hidelinks,breaklinks]{hyperref}

  {\vskip 12pt\noindent{\large\textbf{Nomenclature}}\\[6pt]%
   \begin{tabular}{@{}ll}}%
  {\end{tabular}\vskip 12pt}

\newcommand{\tablenotes}[1]{\vskip 4pt\noindent{\footnotesize #1}}

\providecommand{\keywords}[1]{\vskip 8pt\noindent\textbf{Keywords:} #1}

\title{\bfseries Physics-Informed Neural Networks for Depth-Averaged\\
Granular Avalanche Dynamics on Curved Topography}

\author[1]{Pujan Pranavkumar Purohit}
\author[1]{Pradyumn Singh Sikarwar}
\author[1]{Vishal Sharma}
\author[1,*]{Gaurav Bhutani}
\affil[1]{School of Mechanical and Materials Engineering,
Indian Institute of Technology Mandi, Himachal Pradesh 175075, India}
\affil[*]{Corresponding author. \texttt{gaurav@iitmandi.ac.in}}

\date{}

\begin{document}

\maketitle

\begin{abstract}
Physics-informed neural networks (PINNs) provide a mesh-free framework for solving governing equations, but their application to granular avalanche dynamics over curved terrain remains largely unexplored. This study extends a depth-averaged PINN formulation based on the Savage--Hutter equations to an exponentially curved chute with spatially varying inclination and a strain-rate-dependent Mohr--Coulomb earth-pressure closure. The model is validated against measured front- and rear-edge trajectories from a laboratory granular-avalanche experiment, with selected observations withheld from training. A staged temporal curriculum proved essential for accurate prediction, reducing the held-out trajectory error by approximately two orders of magnitude compared with training over the full time domain from the outset. Sparse-data experiments further showed that observation placement was more influential than observation number within the configurations tested. Four observations bracketing the transition from acceleration to deceleration achieved nearly the same accuracy as the eight-observation reference configuration, whereas observations clustered at early or late times performed poorly. The results demonstrate the importance of both training strategy and informative data placement when applying PINNs to granular flows over curved topography.
\end{abstract}


\vskip 12pt

\section{Introduction}

Rapid granular avalanches of snow, rock, and debris rank among the most destructive hazards affecting mountain communities and infrastructure. Their behaviour is notoriously difficult to characterise empirically. These events are sudden and rarely repeatable under controlled conditions, leaving physics-based modelling as the principal means of anticipating runout distances and impact zones. Because the flow thickness of such avalanches is typically far smaller than the distance they travel downslope, the vertical dependence of the governing equations can be integrated out, reducing a three-dimensional problem to a depth-averaged system for the flow height and the depth-averaged velocity. The formulation of Savage and Hutter \cite{savagehutter1989} remains the most influential model of this class, embedding inertia, basal Coulomb friction, and internal Mohr--Coulomb earth pressure within a compact hyperbolic system of conservation laws.

A substantial body of work has since extended this framework beyond the idealised planar incline on which it was first posed. Curvature of the bed was among the earliest generalisations pursued. Hutter and Koch \cite{hutterkoch1991} conducted a systematic series of experiments on an exponentially curved chute and compared the measured motion against depth-averaged predictions, while Greve and Hutter \cite{grevehutter1993} extended this to convex and concave geometries. Subsequent work generalised the theory of unconfined flow over partly curved surfaces \cite{greve1994}, to complex shallow basal topography via curvilinear reference surfaces \cite{gray1999, wieland1999}, and to curved and twisted channels \cite{pudasaini2003}. Alongside these theoretical developments, robust numerical methods have been established for solving the resulting equations, including shock-capturing and front-tracking schemes \cite{tai2002}, kinetic schemes for Saint-Venant-type systems \cite{mangeney2003}, and the parallel adaptive-mesh solver \texttt{TITAN2D} \cite{pitman2003,patra2005}, which is now widely used for operational hazard assessment. Accurate as these solvers are, deploying them is far more effortful. Mesh generation over irregular terrain, careful treatments of the migrating wet--dry front, and codes written to handle specific terms in governing equations---all impose considerable overhead before a single simulation can be run.

Physics-informed neural networks (PINNs) offer an appealing alternative to this discretisation-centred workflow \cite{raissi2019, karniadakis2021}. Rather than partitioning the domain into cells, PINNs encode the governing equations, initial and boundary conditions and any available observations directly into a neural network's training objective, with the required derivatives obtained through automatic differentiation \cite{baydin2018}. The approach has been applied productively to depth-averaged flow problems in hydraulic settings, including shallow-water flow on a sphere \cite{bihlo2022},  benchmarking against finite-volume solvers for free-surface flow \cite{qi2024}, downscaling of large-scale river models \cite{feng2023}, and unsteady riverine and complex-terrain domains \cite{yin2025,tian2025}. 

Sikarwar et al. \cite{sikarwar2026} addressed this gap directly, developing a PINN framework for the depth-averaged Savage--Hutter equations and validating it in two stages: a one-dimensional formulation in which height and velocity were learned jointly and verified against the analytical similarity solution, and a two-dimensional formulation validated against laboratory pile-collapse measurements \cite{maeno2013} with TITAN2D providing numerical comparison.
Both stages of their study were, however, restricted to a planar inclined bed with a constant earth-pressure coefficient, and natural terrain seldom conforms to either idealisation. Once the bed curves, the local inclination becomes a function of downslope position, and the flow alternatively extends and compresses as it negotiates regions of differing slope. Under these conditions, a fixed earth-pressure coefficient is no longer defensible: the closure must instead alternate between the active and passive Mohr--Coulomb states according to the local strain rate, introducing a nonlinear constitutive coupling between the depth-averaged velocity gradient and the earth-pressure term that was absent when a constant earth-pressure coefficient was used in the planar formulation. Curvature imposes a second, equally consequential difficulty. Once the bed angle varies with position, the governing system admits no closed-form solution, so an analytical benchmark that anchored verification in the planar case is unavailable, and validation must instead rest on measured experimental trajectories.

These complications bear directly on trainability, which remains the principal obstacle to the practical deployment of PINNs. Considerable effort has focused on diagnosing and mitigating the pathologies that arise, including imbalanced gradient flow between loss terms \cite{wangteng2021}, spectral bias and its consequences for convergence \cite{wangyu2022}, and violation of temporal causality during optimisation \cite{wang2024causality}. Among the remedies proposed, curriculum strategies, in which the network is first exposed to a simpler or temporally restricted version of the problem before the full domain is introduced \cite{bengio2009, krishnapriyan2021}, have proved effective across a range of stiff and strongly nonlinear systems. A second and considerably less examined question concerns the observational data itself. Where sparse measurements are used to supplement the physics constraint, most studies treat the quantity of data as the governing variable, and comparatively little attention has been paid to where in the domain those measurements are most informative, although recent work on sensor placement for inverse problems suggests the effect may be substantial \cite{venianakis2026}.

This paper extends the depth-averaged PINN framework to the curved setting described above, and examines both questions in that context. Flow height and depth-averaged velocity are learned jointly on a chute with spatially varying inclination and a strain-rate-dependent earth-pressure closure, and the resulting predictions are validated against measured front and rear edge trajectories from the curved-chute experiments of Hutter and Koch \cite{hutterkoch1991}, with a subset of the recorded time instances withheld entirely from training. Against this benchmark, we address two questions: whether staged curriculum training is essential for accurate trajectory prediction in this coupled setting, and whether the temporal placement of the sparse observations used to guide the network influences held-out accuracy as strongly as their number. The remainder of the paper is organised as follows. Section~\ref{sec:methodology} sets out the governing equations, chute geometry, and the neural network architecture. Section~\ref{sec:results} describes experimental validation, followed by the results, including curriculum and data-placement ablations. Section~\ref{sec:conclusions} presents the study's conclusions and highlights key limitations.

\section{Methodology}
\label{sec:methodology}

\subsection{Governing equations}
\label{sec:governing}

The depth-averaged Savage--Hutter system solved here retains the structure established for the planar case \cite{sikarwar2026}, but generalises it in two respects: the bed inclination angle becomes a function of downslope position, and the earth-pressure coefficient is allowed to vary with the local longitudinal strain rate.

Following Savage and Hutter \cite{savagehutter1989}, the characteristic flow thickness $H$ is distinguished from the downslope length scale $L$, defining the shallow-flow aspect ratio
\begin{equation}
\epsilon=\frac{H}{L}\ll1.
\label{eq:aspect}
\end{equation}
Flow height is scaled by $H$, downslope distance by $L$, velocity by $\sqrt{gL}$, and time by $\sqrt{L/g}$. All quantities appearing in the governing equations below are therefore non-dimensional. Because the vertical and downslope coordinates are scaled by different characteristic lengths, the aspect ratio $\epsilon$ remains explicitly in the depth-averaged momentum equation through the coefficient $\beta_0$. The assumption $\epsilon\ll1$ underlies the shallow-layer reduction from the three-dimensional granular-flow equations to the depth-averaged system considered here.

Mass conservation is unaffected by either generalisation and retains the planar form,
\begin{equation}
\frac{\partial h}{\partial t}
+\frac{\partial(hu)}{\partial x}=0,
\label{eq:continuity}
\end{equation}
where $h(x,t)$ and $u(x,t)$ denote the non-dimensional flow height and depth-averaged tangential velocity, respectively. Here, $x$ denotes the arc-length coordinate measured along the curved chute.

The depth-averaged momentum balance retains the principal contributions of the planar Savage--Hutter formulation, while allowing the local bed inclination to vary with the downslope coordinate \cite{hutterkoch1991}:
\begin{equation}
\frac{\partial u}{\partial t} + u\frac{\partial u}{\partial x}
= \sin\zeta(x) - \tan\delta\cos\zeta(x) - \beta_0 \frac{\partial h}{\partial x},
\label{eq:momentum}
\end{equation}
where $\delta$ is the bed friction angle and
\begin{equation}
\beta_0 = \epsilon k(u_x)\cos\zeta(x),
\label{eq:beta}
\end{equation}
with $\epsilon$ the aspect ratio and $k$ the earth-pressure coefficient.
The three terms on the right-hand side of Equation~(\ref{eq:momentum}) represent, respectively, the downslope component of gravity, basal Coulomb friction, and the depth-averaged earth-pressure gradient. The basal friction term represents the shear resistance exerted by the bed through the Coulomb model, whereas the earth-pressure term accounts for the effect of the internal granular stress state through the coefficient $k$. 
Internal shear stresses are neglected under the approximation adopted here, as their contribution is assumed to be small relative to the dominant basal shear resistance in dense granular avalanche flows. Likewise, stresses and velocity variations across the flow depth are not explicitly resolved, having been eliminated through the shallow-layer depth averaging underlying the Savage--Hutter model. The full derivation of the depth-averaged system, together with the treatment of the basal boundary conditions and the curvature terms arising in bed-fitted coordinates, is given in  Sikarwar et al. \cite{sikarwar2026} and in Sharma and Bhutani \cite{sharma2026computational} and is not reproduced here.

The spatial dependence of $\zeta$ modifies both the gravitational forcing and basal friction. Consequently, the net downslope contribution $\sin\zeta-\tan\delta\cos\zeta$ changes sign when the local bed angle falls below the basal friction angle. This transition, absent on a planar bed of fixed inclination that remains above the friction angle, promotes the deceleration and eventual arrest of the flow and is central to the results reported in Section~\ref{sec:results}.

The second generalisation concerns the earth-pressure closure. As the granular mass moves over the curved chute, different portions of the flow may undergo local extension or compression, and the closure must respond accordingly. The coefficient is therefore allowed to switch between the active and passive Mohr--Coulomb states, $k_{\mathrm{act}}$ and $k_{\mathrm{pass}}$, according to the sign of the local strain rate:
\begin{equation}
k(u_x) = \sigma(5u_x)\,k_{\mathrm{act}}
+ \bigl(1-\sigma(5u_x)\bigr)\,k_{\mathrm{pass}},
\label{eq:kswitch}
\end{equation}
where $\sigma$ denotes the logistic sigmoid and $u_x = \partial u/\partial x$ is the local strain rate. Equation (\ref{eq:kswitch}) follows the standard convention: an extending flow ($u_x > 0$) yields at the lower active pressure, whereas a compressing flow ($u_x < 0$) resists at the higher passive pressure, with $k$ blending smoothly between the two limits.

The sharpness constant of $5$ appearing in Equation (\ref{eq:kswitch}) was chosen deliberately low, so as to produce a wide, smooth transition about $u_x = 0$ rather than an abrupt switch. The motivation is numerical rather than physical. Because the switch is evaluated on the network's own estimate of $u_x$, which is necessarily poor during early training, a sharp switch would cause $k$ to alternate discontinuously between the active and passive states on essentially every sign fluctuation of a quantity the network has not yet learned to represent. The smoother blend trades some physical sharpness in the Mohr--Coulomb switch for a training signal that varies continuously with the network parameters.

The coupling is particularly important for the trainability of the PINN. In Equations (\ref{eq:momentum})--(\ref{eq:kswitch}), the earth pressure coefficient depends on the local gradient of the velocity field predicted by the network. Consequently, errors in $u$ directly affect the constitutive closure through $u_x$, which in turn modifies the momentum residual used to update the same network parameters. This creates a feedback loop between the network prediction, the earth-pressure closure, and the governing-equation residual that is absent in the planar formulation. The staged training strategy described in Section~\ref{sec:architecture} was therefore introduced to progressively establish a stable velocity field before imposing the full coupled dynamics.

The parameter values adopted throughout are those reported for Experiment 73 of Hutter and Koch \cite{hutterkoch1991}: a bed friction angle $\delta = 29^\circ$, and an aspect ratio $\epsilon = 0.1$. The corresponding limiting earth-pressure coefficients are $k_{\mathrm{act}} = 1.2988$ and $k_{\mathrm{pass}} = 5.5867$.

\subsection{Chute geometry}
\label{sec:geometry}

The bed geometry follows the exponentially curved chute of Hutter and Koch \cite{hutterkoch1991}, in which a rough curved surface descends from a steep upper section, through a progressively flattening transition, to a shallow runout region where the flow decelerates and comes to rest (See Figure~\ref{fig:curved_a8}). 
The local bed angle varies with downslope position according to
\begin{equation}
\zeta(x) = 60^\circ\, e^{-0.1x},
\label{eq:zeta}
\end{equation}
over the spatial domain $x \in [0, 22]$, with the rigid back wall of the chute located at $x = 0$. The simulated time
interval spans $t \in [0, T_{\max}]$ with $T_{\max} = 14$, chosen so that the granular mass remains within the domain and reaches its final deposit configuration before the end of the window.

The granular mass is released from rest, so that $u(x,0) = 0$ throughout, with the initial pile profile prescribed as
\begin{equation}
h(x,0) = h_0(x) = 1 - \left(\frac{x-1.25}{1.25}\right)^{2}, \qquad x \in [0,\,2.5],
\label{eq:ic}
\end{equation}
and $h_0(x) = 0$ outside this interval, corresponding to a parabolic pile of unit peak height centred at $x=1.25$.

\subsection{Network Architecture}
\label{sec:architecture}

The solution fields were represented by a single fully connected multilayer perceptron mapping the space--time coordinate pair to both depth-averaged quantities,
\begin{equation}
\mathcal{N}_\theta : (x,t) \mapsto (h,u),
\label{eq:mapping}
\end{equation}
with trainable parameters $\theta$. The network comprises six hidden layers of $128$ neurons each with hyperbolic tangent activations, a choice that is standard for PINNs applied to conservation laws \cite{raissi2019} and that provides smooth higher-order
derivatives required to evaluate the momentum residual through automatic differentiation \cite{baydin2018}. Input coordinates were linearly normalised to $[-1,1]$ using the domain bounds given in Section~\ref{sec:geometry} before entering the network.

The network mapped the space--time coordinates $(x,t)$ simultaneously to the two physical fields, $h(x,t)$ and $u(x,t)$, using a common set of hidden layers. The flow height is
passed through a squared softplus activation with sharpness parameter $\beta=3$,
\begin{equation}
h(x,t) = \bigl[\mathrm{softplus}_{\beta=3}(z_h)\bigr]^{2}, \qquad
\mathrm{softplus}_{\beta}(z) = \frac{1}{\beta}\ln\!\bigl(1+e^{\beta z}\bigr),
\label{eq:softplus}
\end{equation}
which enforces $h \geq 0$ by construction and thereby prevents the network from representing physically meaningless negative depths in the dry region ahead of and behind the pile. The velocity output was left unconstrained, since $u$ may legitimately take either sign as the mass extends and compresses. Weights were initialised using the Xavier uniform scheme, with biases set to zero.

The network was trained by minimising a composite objective combining the governing physics, the prescribed initial and boundary conditions, a small set of experimental observations, and a global conservation constraint:
\begin{equation}
\mathcal{L}_{\mathrm{total}} =
w_{\mathrm{pde}}\mathcal{L}_{\mathrm{PDE}} +
w_{\mathrm{ic}}\mathcal{L}_{\mathrm{IC}} +
w_{\mathrm{wall}}\mathcal{L}_{\mathrm{wall}} +
w_{\mathrm{data}}\mathcal{L}_{\mathrm{data}} +
w_{\mathrm{mass}}\mathcal{L}_{\mathrm{mass}}.
\label{eq:loss}
\end{equation}

The residual loss $\mathcal{L}_{\mathrm{PDE}}$ enforces both Equation (\ref{eq:continuity}) and Equation (\ref{eq:momentum}) at collocation points distributed over the spatio-temporal domain, with the continuity and momentum residuals combined using fixed weights of $5.0$ and $25.0$, respectively. 
The momentum residual is additionally masked to zero wherever the predicted flow height falls below $h=0.01$, so the network is not penalised for enforcing momentum balance in regions effectively devoid of granular material. This treatment is particularly relevant to the strain-rate-dependent closure adopted here, since the local velocity and its gradient $u_x$ have no meaningful physical interpretation in dry regions. Without masking, these unconstrained velocity gradients could still influence the active--passive earth-pressure coefficient through Equation~(\ref{eq:kswitch}) and spuriously contribute to the momentum residual.

The initial-condition loss $\mathcal{L}_{\mathrm{IC}}$ measures the mismatch between the network prediction at $t=0$ and the prescribed state of Equation (\ref{eq:ic}), with the height and velocity components weighted $100.0$ and $50.0$ respectively. The boundary
loss $\mathcal{L}_{\mathrm{wall}}$ enforces zero flux, $hu = 0$, at the rigid back wall of the chute at $x = 0$. 

The data loss $\mathcal{L}_{\mathrm{data}}$ constrains the predicted height to near-zero at the experimentally measured front and rear edge locations, evaluated at the timestamp of each training row. It is therefore a zero-crossing constraint on where the pile begins and
ends, rather than a fit to its interior shape, which is appropriate given that the experiment records edge positions alone. The auxiliary term $\mathcal{L}_{\mathrm{mass}}$ penalises departure of the implied total mass $\int h \, \mathrm{d}x$ from its initial value, providing a global check on conservation that complements the pointwise residual
enforced through $\mathcal{L}_{\mathrm{PDE}}$.

The loss weights were set to $w_{\mathrm{wall}} = 5.0$, $w_{\mathrm{data}} = 2.0$, and $w_{\mathrm{mass}} = 1.0$. All weights, including the sub-term weights quoted above, were fixed a priori and held constant throughout the study rather than tuned adaptively. Two scaling factors were additionally applied. The PDE residual terms were scaled by $1 + 9(t/T_{\max})$, increasing linearly with time, and the initial-condition terms by $\max\!\left(0.2,\; 1 - 0.7\, t_{\mathrm{win}}/T_{\max}\right)$, decaying as the curriculum window $t_{\mathrm{win}}$ widens. Together, these shift the objective's emphasis from reproducing the initial state to satisfying the governing equations at later times as training progresses.

Training proceeded in four sequential phases, each addressing a specific difficulty introduced by the coupled, curved-chute setting. In the first phase, the network was trained for $3{,}000$ Adam steps to reproduce the prescribed initial state alone, $h(x,0) = h_0(x)$ and $u(x,0) = 0$, before introducing any physics or data loss. This places the parameters near a physically sensible starting point rather than requiring the optimiser to satisfy the governing equations from a random initialisation.

The second phase introduced the PDE residual and data losses under a staged temporal curriculum. Rather than exposing the network to the full simulation window at once, these losses were enforced only up to a training time window $t_{\mathrm{win}}$, which was widened in eight steps: $t_{\mathrm{win}} = 0.5$, $1.0$, $2.0$, $4.0$, $6.0$, $9.0$,
$12.0$ and $14.0$, over $37{,}000$ Adam steps in total. Exposing the dynamics progressively in time encourages the network to resolve early-time behaviour accurately before later times are introduced, following the broader principle that ordering training from simpler to harder sub-problems stabilises optimisation \cite{bengio2009,krishnapriyan2021} and respects the temporal causal structure of the underlying evolution \cite{wang2024causality}.

In the third phase, once the curriculum reached the full time window, we took a further $8{,}000$ Adam steps over the entire domain with a cosine-annealed learning rate, allowing the solution to be refined jointly across all times rather than stage by stage. The fourth and final phase applied a single L-BFGS optimisation with a strong-Wolfe line search and history size $50$, run for up to $400$ iterations on one fixed collocation batch sampled at $t = T_{\max}$.
Each training step drew $12{,}000$ collocation points for the PDE residual, $4{,}000$ for the initial condition, and $500$ for the wall boundary.

One feature of this schedule deserves particular emphasis, since it underlies the data-placement results reported in Section~\ref{sec:results}. The data loss is itself gated by the curriculum from the second phase onward: an experimental row contributes to $\mathcal{L}_{\mathrm{data}}$ only once $t_{\mathrm{win}}$ has reached its own timestamp. Observations at early times therefore influence the optimisation across most of the curriculum, whereas those at late times enter only during the final one or two stages. Which rows are selected for training consequently determines not only what information is supplied to the network, but also how long that information is available to shape the solution.

\subsection{Edge extraction and evaluation metric}
\label{sec:metric}

Validation was based on the front and rear edge positions reported by Hutter and Koch \cite{hutterkoch1991}, as Experiment 73 provides no measurements of the interior height or velocity fields. Since the network predicts a continuous height field $h(x,t)$ whose softplus output does not vanish at a unique edge, the front and rear positions were extracted using a prescribed depth threshold: the predicted height field was sampled on a uniform grid of $1200$ points across the spatial domain at each evaluation time, and the front/rear edges were taken as the outermost points where $h > 0.025$.

The predicted edge trajectories were compared with the experimental measurements using the mean squared edge error,
\begin{equation}
\text{track\_err}
=
\frac{1}{N}
\sum_{i=1}^{N}
\left[
\Delta f(t_i)^2+\Delta r(t_i)^2
\right],
\label{eq:trackerr}
\end{equation}
where
\begin{equation}
\Delta f(t_i)
=
x_{f,\mathrm{pred}}(t_i)-x_{f,\mathrm{meas}}(t_i),
\qquad
\Delta r(t_i)
=
x_{r,\mathrm{pred}}(t_i)-x_{r,\mathrm{meas}}(t_i)
\label{eq:edgeerrors}
\end{equation}
denote the front- and rear-edge errors, respectively.

Experiment 73 contains sixteen time-stamped observations, of which four ($t\approx1.7$, $5.1$, $8.5$ and $11.8$) were withheld entirely from training and used for evaluation, giving $N=4$. Thus, $\text{track\_err}$ is a held-out generalisation metric rather than a training residual. The withheld observations were distributed across the trajectory to represent different stages of the motion.

\section{Results}
\label{sec:results}

The results are presented in three parts, focusing respectively on validation against the experimental trajectories, the effect of the staged training strategy, and the influence of sparse-data placement. Throughout, $\text{track\_err}$ is evaluated exclusively on the four withheld observations, while the mass-conservation error is reported alongside it to assess conservation independently of trajectory accuracy.

\subsection{Validation against experimental trajectories}
\begin{figure}[!htbp]
\centering
\includegraphics[width=0.85\linewidth]{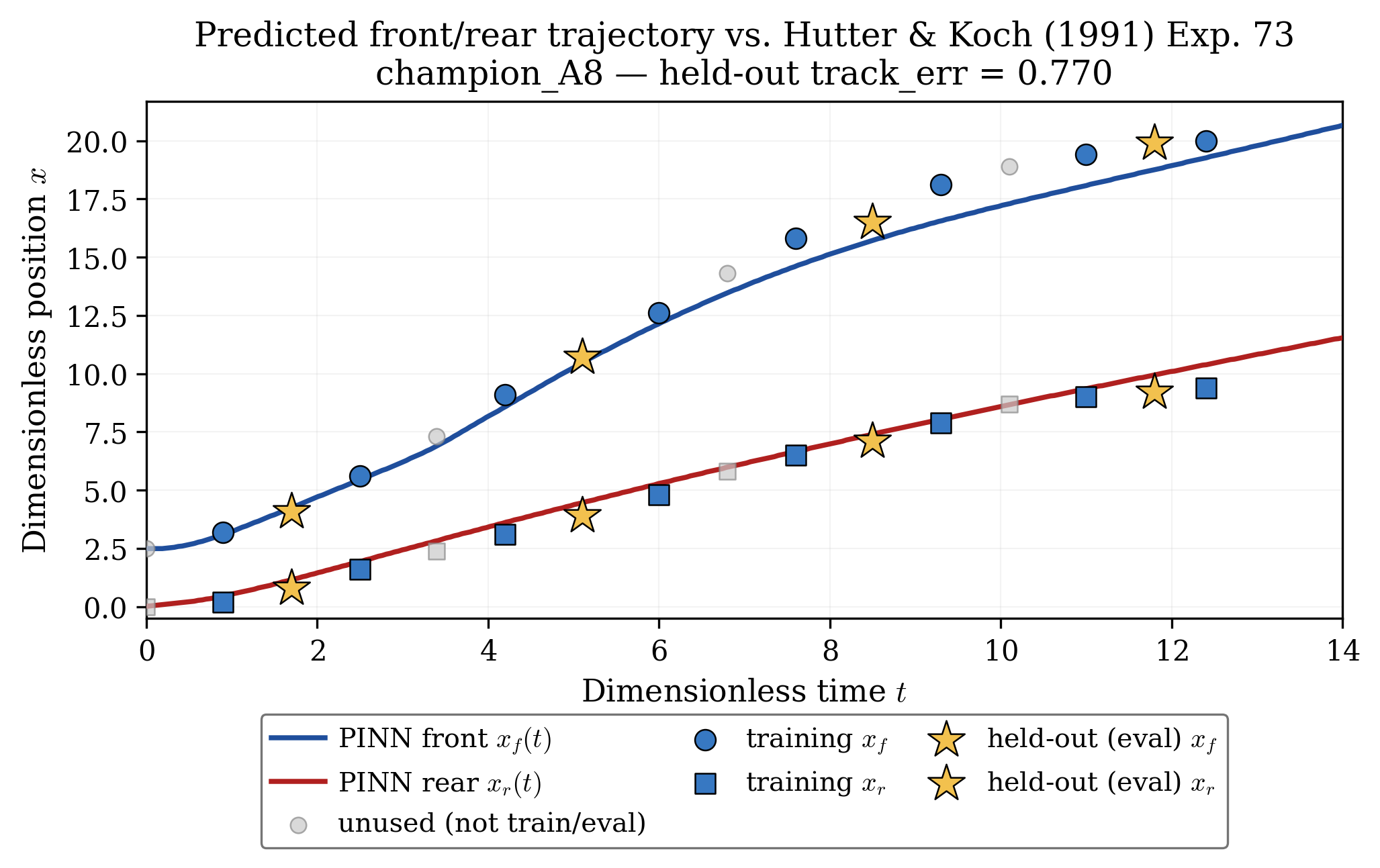}
\caption{Predicted front and rear edge trajectories for the reference
configuration (eight training rows) against Experiment 73 of Hutter
and Koch \cite{hutterkoch1991}. Stars mark the four held-out rows,
which were withheld entirely from training. Held-out
$\text{track\_err} = 0.770$.}
\label{fig:headline_a8}
\end{figure}

The reference configuration, trained using eight of the twelve available training observations, reproduced the measured front and rear edge trajectories over the full simulation window and achieved a held-out $\text{track\_err}$ of $0.77$ (Figure~\ref{fig:headline_a8}). The corresponding held-out RMSE was $0.71$ at the front edge and $0.52$ at the rear edge, with a maximum absolute error of $1.15$. The somewhat larger error at the front edge is consistent with the nature of the flow: as the granular mass spreads, the leading edge develops a thin, low-depth tail, making its detected position more sensitive to the chosen extraction threshold than the relatively well-defined trailing edge.

\begin{figure}[!htbp]
\centering
\includegraphics[width=0.85\linewidth]{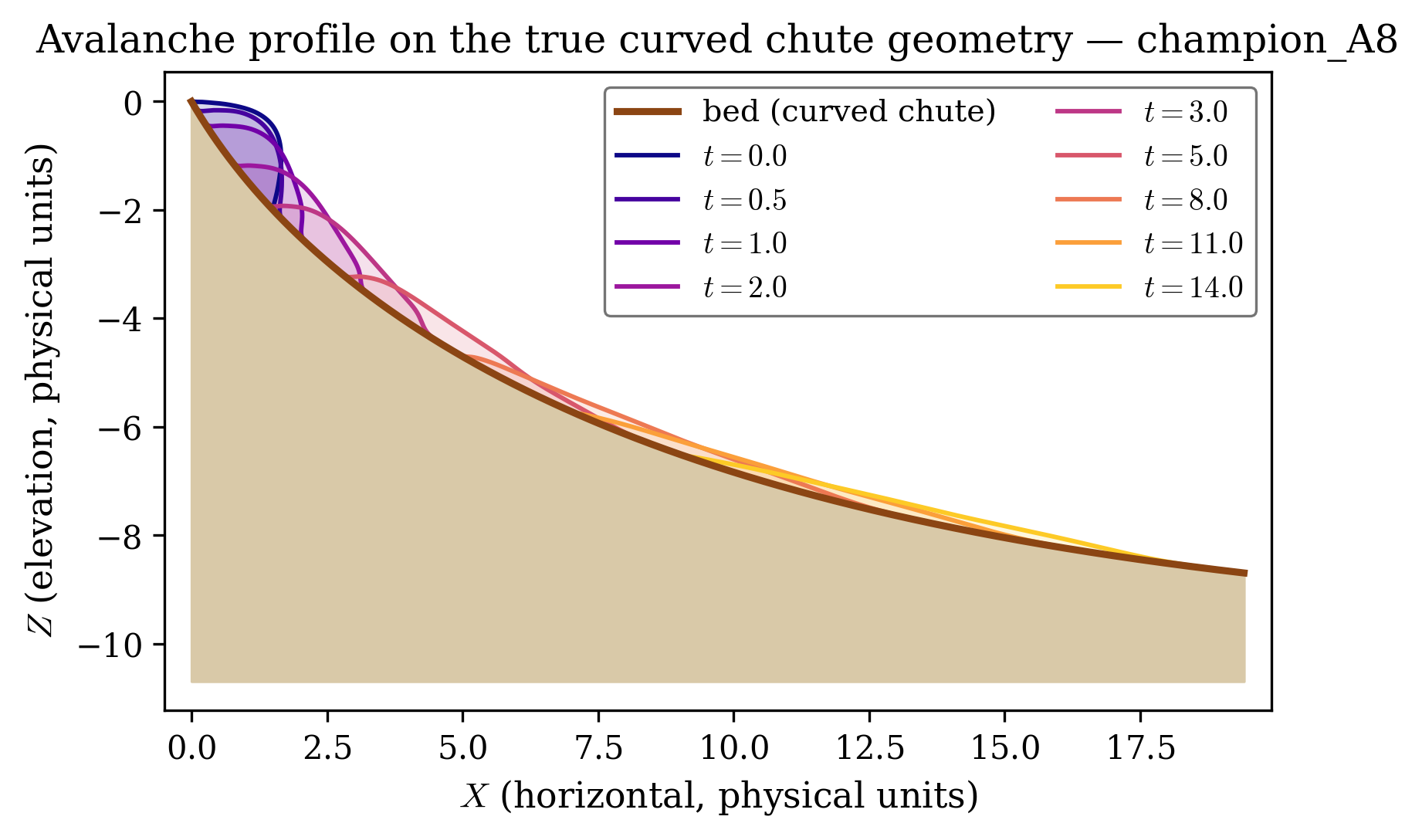}
\caption{Predicted height field on the curved chute geometry for the
reference configuration, shown at $t = 0$, $0.5$, $1$, $2$, $3$, $5$,
$8$, $11$ and $14$. The bed profile is overlaid for reference.}
\label{fig:curved_a8}
\end{figure}

The predicted height field on the curved chute is shown in Figure~\ref{fig:curved_a8}. The granular mass remained coherent throughout the simulation, including across the transition region at $x\approx5$--$9$, where the local bed angle falls below the friction angle. No spurious fragmentation or negative-depth artefacts were observed. Since the experimental data contain only front and rear edge positions, the interior height and velocity fields are model predictions and cannot be independently validated against the experiment. They are therefore presented to illustrate the qualitative predicted flow structure rather than to directly assess predictive accuracy.

\begin{figure}[!htbp]
\centering
\includegraphics[width=0.85\linewidth]{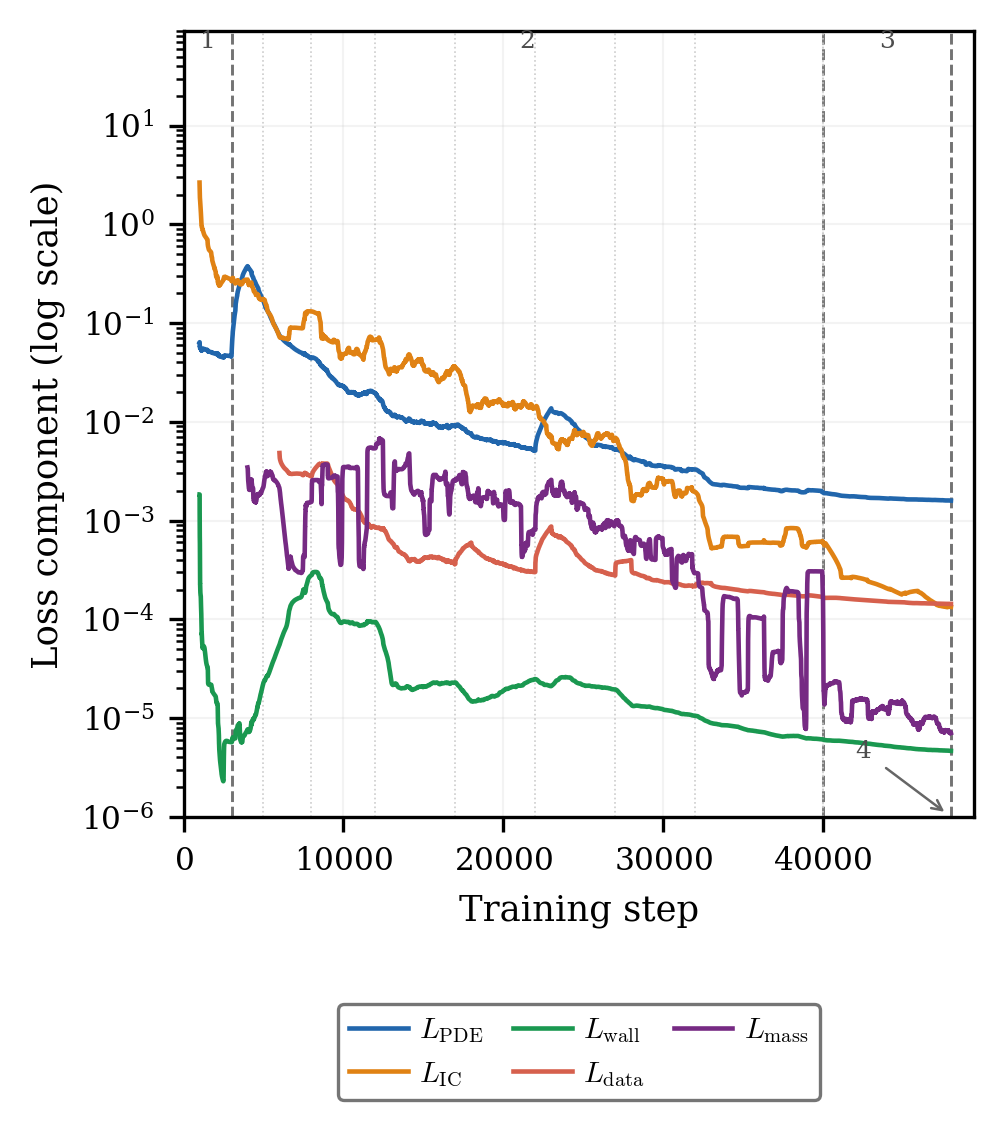}
\caption{Evolution of the five loss components ($\mathcal{L}_{\mathrm{PDE}}$,
$\mathcal{L}_{\mathrm{IC}}$, $\mathcal{L}_{\mathrm{wall}}$,
$\mathcal{L}_{\mathrm{data}}$, $\mathcal{L}_{\mathrm{mass}}$) against
training step (log scale) for the reference configuration. Vertical
dashed lines mark the boundaries between the four training phases
described in Section~\ref{sec:architecture}, numbered sequentially; the
arrow indicates the brief final L-BFGS phase at the end of training.}
\label{fig:loss_components}
\end{figure}

\subsection{Effect of the staged curriculum}

The contribution of the staged curriculum was isolated using three configurations with the same eight-row training budget, network architecture and total number of optimisation steps, differing only in the curriculum schedule (Table~\ref{tab:curriculum}). Removing the staging entirely and applying the PDE and data losses over the full time interval from the outset increased $\text{track\_err}$ from $0.77$ to $69.03$. A coarser three-stage curriculum performed worse, with an error of $120.33$. The two loss-scaling factors produced improvements in the same direction, although their effects were less pronounced (Table~\ref{tab:scaling}). Removing the time-dependent PDE weight increased $\text{track\_err}$ to $2.56$, removing the initial-condition decay increased it to $1.20$, and removing both factors resulted in $2.85$. Across all configurations in the two ablations, the mass-conservation error varied only from $0.12$ to $0.20\,\%$.

\begin{table}[!t]
\caption{Curriculum ablation. All three configurations use the same eight-row training budget, the same four held-out rows, and a matched total step count; only the curriculum schedule differs. Mass-conservation error is essentially unaffected by the choice of schedule, as expected of a global integral, while the trajectory metric degrades by roughly two orders of magnitude without staging.}
\label{tab:curriculum}
\centering
\begin{tabular}{|l|c|c|}
\hline
\textbf{Configuration} & \textbf{track\_err (held-out)} & \textbf{Mass-conservation error}\\
\hline
Eight-stage curriculum (adopted) & 0.77 & 0.16\,\%\\
No staging (single window) & 69.03 & 0.12\,\%\\
Coarse three-stage curriculum & 120.33 & 0.20\,\%\\
\hline
\end{tabular}
\end{table}

\begin{table}[!t]
\caption{Loss-scaling ablation, on the same eight-row training budget
and full training schedule as Table~\ref{tab:curriculum} and otherwise
identical settings. The reference configuration applies both the
time-increasing PDE weight and the curriculum-window decay on the
initial-condition terms; each ablation removes one or both.}
\label{tab:scaling}
\centering
\begin{tabular}{|l|c|c|}
\hline
\textbf{Configuration} & \textbf{track\_err (held-out)} & \textbf{Mass-conservation error}\\
\hline
Both scaling factors (adopted) & 0.77 & 0.16\,\%\\
No PDE time-weight & 2.56 & 0.12\,\%\\
No IC decay & 1.20 & 0.13\,\%\\
Neither factor & 2.85 & 0.18\,\%\\
\hline
\end{tabular}
\end{table}

The results indicate that staging was essential for accurate prediction on the curved geometry rather than being merely a training refinement. The underlying reason is the strain-rate-dependent constitutive closure discussed in Section~\ref{sec:governing}. In the planar formulation, the momentum residual contains no equivalent feedback through the earth-pressure coefficient, and the companion study \cite{sikarwar2026} did not require a curriculum. In the present formulation, however, Eqs.~(\ref{eq:momentum})--(\ref{eq:kswitch}) make the earth-pressure term dependent on $\partial u/\partial x$. An inaccurate velocity prediction therefore produces an inaccurate strain rate, which in turn modifies the earth-pressure coefficient and hence the momentum residual used to correct the velocity. This feedback makes the optimisation considerably more sensitive during the early stages of training. The curriculum mitigated this difficulty by progressively expanding the temporal domain. The network first resolved the early motion, where the solution remained close to the prescribed initial state and the flow was nearly at rest, before being exposed to the later transition and runout dynamics. This interpretation is consistent with previous observations that respecting temporal causality can improve PINN trainability \cite{wang2024causality} and that progressively increasing problem difficulty can stabilise optimisation for stiff learning problems \cite{bengio2009,krishnapriyan2021}.

The weak sensitivity of the mass-conservation error to these training changes is also important. Configurations with trajectory errors differing by nearly two orders of magnitude nevertheless conserved mass to within approximately $0.2\,\%$. A global integral constraint controls the total amount of material represented by the solution but does not constrain its spatial distribution. Consequently, mass conservation alone cannot serve as a reliable accuracy diagnostic for runout problems in which the location of the granular mass is the quantity of interest. A held-out spatial metric such as $\text{track\_err}$ is therefore required.

\subsection{Effect of sparse-data placement}

The held-out accuracy was largely insensitive to the number of evenly spaced training observations between eleven and six rows, with $\text{track\_err}$ varying only from $0.77$ to $0.97$ (Table~\ref{tab:budget}). Reducing the training set from six to four rows, however, resulted in an abrupt deterioration rather than a gradual loss of accuracy, with $\text{track\_err}$ reaching $65.08$.

\begin{table}[!t]
\caption{Training-data budget against held-out accuracy, using rows spread evenly across the record at each step, ordered from most to fewest. Accuracy remains nearly unchanged from eleven to six rows but deteriorates sharply between six and four rows, while the mass-conservation error provides no indication of the failure.}
\label{tab:budget}
\centering
\begin{tabular}{|c|c|c|c|c|}
\hline
\textbf{Training rows (of 16)} & \textbf{Held-out rows} &
\textbf{Unused rows} & \textbf{track\_err} &
\textbf{Mass-cons.\ error}\\
\hline
11 & 4 & 1 & 0.85 & 0.14\,\%\\
10 & 4 & 2 & 0.85 & 0.16\,\%\\
8 (reference) & 4 & 4 & 0.77 & 0.16\,\%\\
6 & 4 & 6 & 0.97 & 0.15\,\%\\
4 (evenly spaced) & 4 & 8 & 65.08 & 0.11\,\%\\
\hline
\end{tabular}
\end{table}

\begin{figure}[!htbp]
\centering
\includegraphics[width=0.85\linewidth]{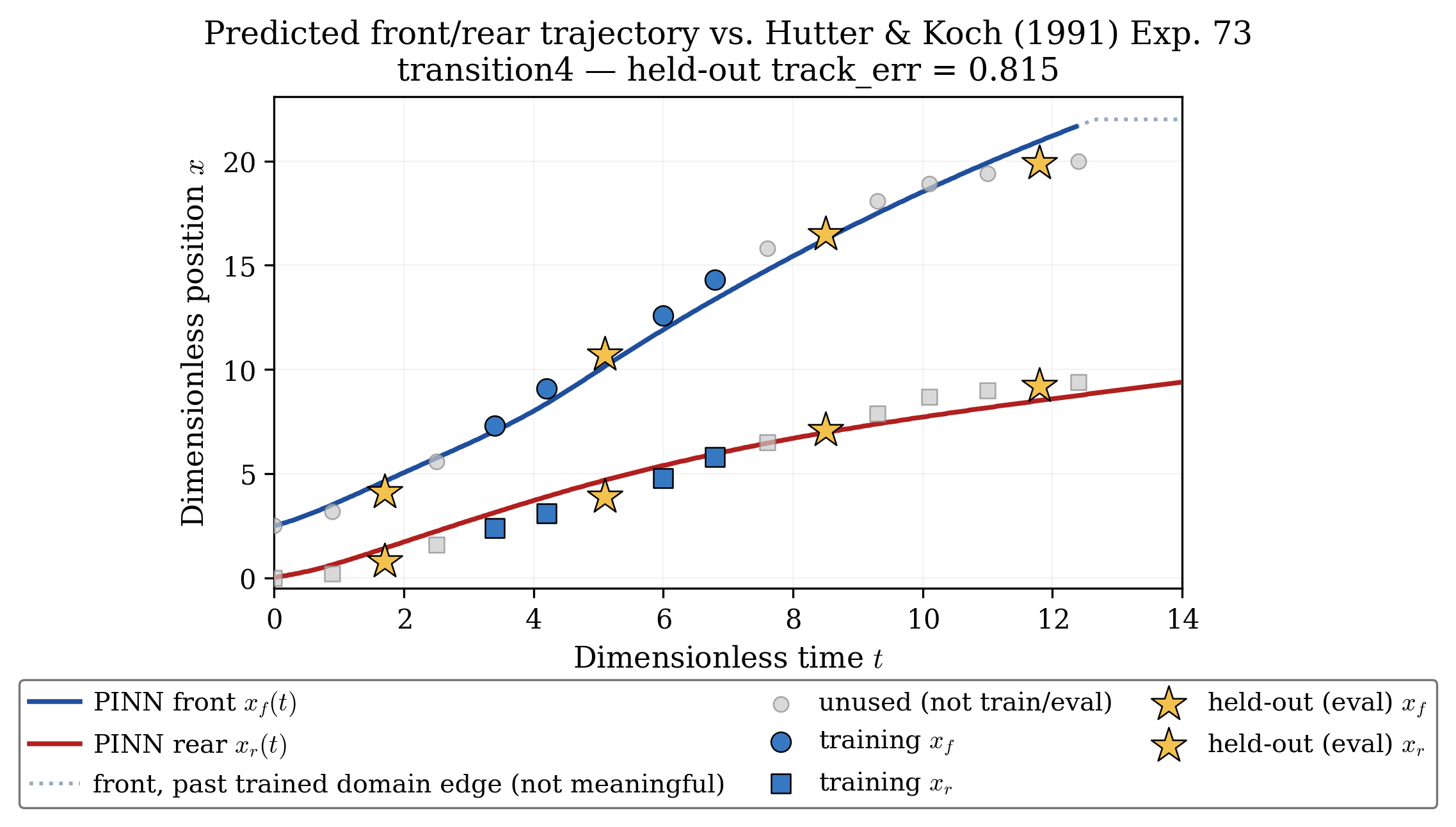}
\caption{Predicted trajectories for the transition-bracketing
configuration using four training rows.  The dotted segment beyond
$t\approx13$ lies outside the evaluation window and reflects domain
truncation rather than physical arrest.}
\label{fig:headline_t4}
\end{figure}

To determine whether four observations were simply insufficient or whether the outcome depended on their placement, three configurations were trained using exactly four observations and differing only in which observations were selected (Table~\ref{tab:placement}). Clustering the observations late in the record produced a $\text{track\_err}$ of $64.18$, while clustering them early gave $42.96$. In contrast, selecting four observations that bracketed the friction-angle transition reduced the error to $0.81$, close to the eight-row reference value of $0.77$ despite using half the training data. The corresponding trajectory prediction is shown in Figure~\ref{fig:headline_t4}, with held-out RMSE values of $0.66$ at the front and $0.61$ at the rear.

\begin{table}[!t]
\caption{Sparse-data placement study. All three configurations use
exactly four training rows and the same four held-out rows as in
Tables~\ref{tab:curriculum}--\ref{tab:budget}; only the placement of
the four training rows differs. The eight-row reference configuration
in Table~\ref{tab:budget} gives $\text{track\_err}=0.77$ for comparison.}
\label{tab:placement}
\centering
\begin{tabular}{|l|c|c|}
\hline
\textbf{Placement} & \textbf{Times used (non-dim.)} &
\textbf{track\_err (held-out)}\\
\hline
Clustered late & 9.3, 10.1, 11.0, 12.4 & 64.18\\
Clustered early & 0.9, 2.5, 3.4, 4.2 & 42.96\\
Transition-bracketing & 3.4, 4.2, 6.0, 6.8 & 0.81\\
\hline
\end{tabular}
\tablenotes{The transition-bracketing value is 0.815 before rounding
to two decimal places.}
\end{table}

The results show that, within the configurations tested, observation placement had a substantially greater influence on held-out accuracy than the number of observations (Table~\ref{tab:placement}). Notably, the dense-early configuration's mean training timestamp lies closer to the transition-bracketing configuration's than the dense-late configuration's does, yet only transition-bracketing matches the reference accuracy — indicating that mean temporal position alone does not account for the placement effect. Two mechanisms may contribute to this behaviour. The transition region is where the net downslope forcing changes sign and the flow begins to decelerate, making observations there particularly informative about the subsequent dynamics. At the same time, these observations are activated relatively early by the curriculum gating described in Section~\ref{sec:architecture}. The transition-bracketing observations at $t=3.4$--$6.8$ therefore influence a substantial portion of the training schedule, whereas the late-clustered observations become active only during the final one or two stages. The present results cannot fully distinguish whether the advantage arises primarily from the physical location of the observations or from the duration over which they influence training. Separating these effects would require activating transition and non-transition observations at the same curriculum stage, or removing the data gating.
Thus, within the configurations examined, where the observations were placed mattered more than how many were supplied. However, further investigation is needed to determine whether the improved performance associated with transition-region observations represents a general feature of the problem.

\begin{figure}[!htbp]
\centering
\includegraphics[width=0.85\linewidth]{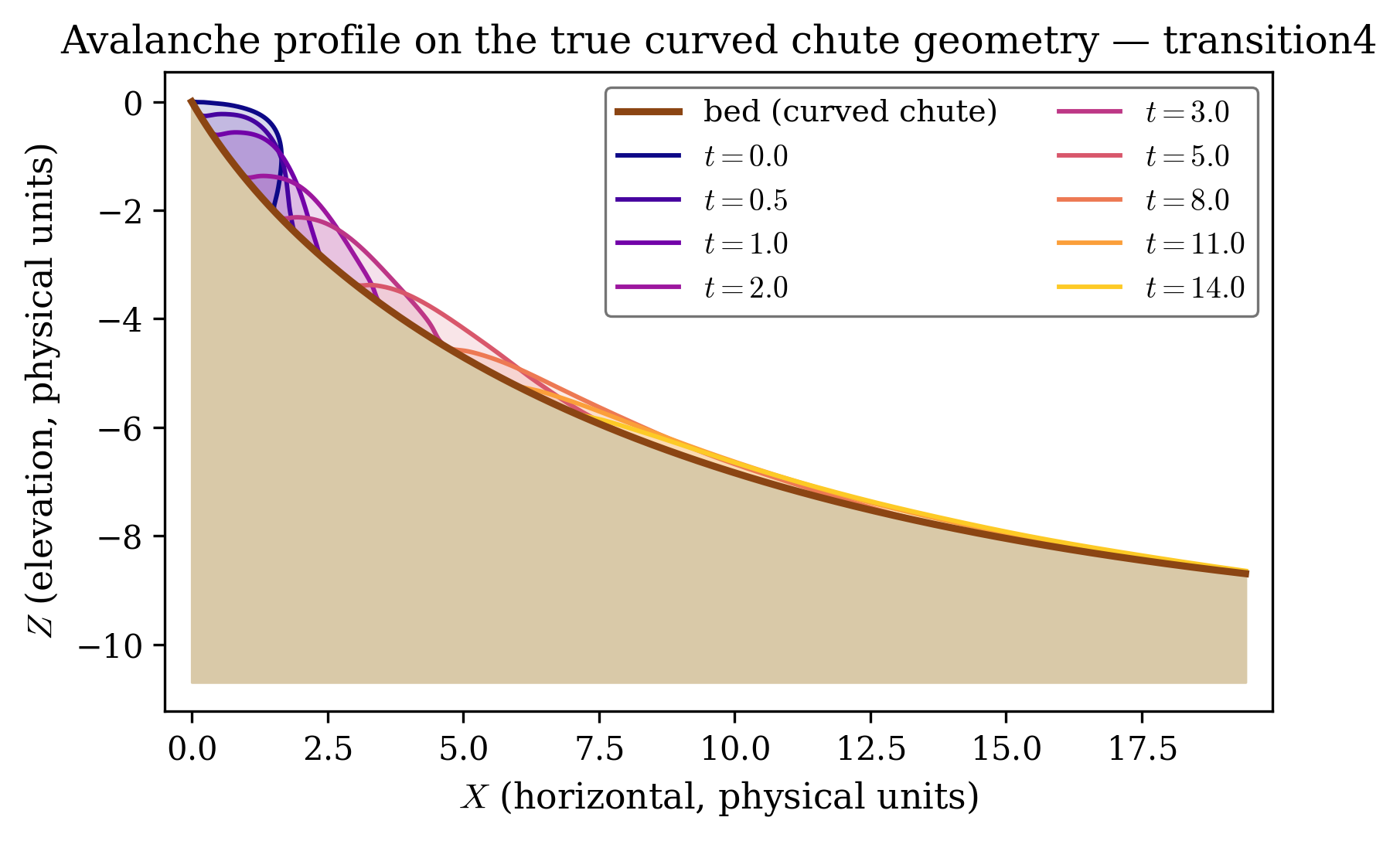}
\caption{Predicted height field for the transition-bracketing
configuration at the same time instants as
Figure~\ref{fig:curved_a8}.}
\label{fig:curved_t4}
\end{figure}

The predicted height field for the transition-bracketing configuration is shown in Figure~\ref{fig:curved_t4}; the corresponding reference-configuration field appears earlier in Figure~\ref{fig:curved_a8}. When checked directly, the two solutions agree closely in the bulk of the flow, with a mean absolute height difference of $0.02$, corresponding to less than $2\,\%$ of the peak depth when averaged over $t=1$--$11$. The derived front position nevertheless shows a larger difference, with an RMS error of $0.81$. The discrepancy is concentrated in the thin leading tail, which contains little mass and therefore contributes negligibly to the bulk height comparison, but can shift the threshold-detected edge appreciably. Thus, the four-observation configuration reproduced the bulk flow of the reference configuration using half the training data, with the remaining discrepancy concentrated in the region where edge extraction is most sensitive.

\section{Conclusions}
\label{sec:conclusions}

A physics-informed neural network framework previously validated for planar inclined beds was extended to a curved chute with a spatially varying bed inclination and a strain-rate-dependent Mohr--Coulomb earth-pressure closure. In the absence of a closed-form solution for the curved geometry, validation was performed against the front and rear edge trajectories measured in Experiment 73 of Hutter and Koch \cite{hutterkoch1991}, with four of the sixteen observations withheld entirely from training. The reference configuration reproduced the measured trajectories with a held-out $\text{track\_err}$ of $0.77$.

Two aspects of the training strategy proved particularly important. First, the temporal curriculum was essential for accurate prediction. Progressive expansion of the training window reduced the held-out error by approximately two orders of magnitude compared with training over the full trajectory from the outset. This behaviour was attributed to the feedback introduced by the strain-rate-dependent closure, in which the earth-pressure term depends on the velocity gradient predicted by the network. Second, the placement of sparse observations had a stronger influence than their number within the configurations tested. Four observations bracketing the friction-angle transition achieved a $\text{track\_err}$ of $0.81$, close to the eight-observation reference value, whereas observations clustered exclusively in the early or late stages produced much larger errors. Mass-conservation error remained below $0.25\,\%$ for all configurations, including those with very poor trajectory predictions. It therefore cannot be used alone as an accuracy criterion when the spatial distribution of the flow is the primary quantity of interest.

Several limitations remain. The experimental dataset contains only sixteen observations from a single experiment, with four used for evaluation, and therefore provides limited evidence for generalisation beyond the present geometry. Validation is restricted to the front and rear edge trajectories because no interior measurements are available; the predicted height and velocity fields should consequently be regarded as model outputs rather than independently validated quantities. Finally, the sparse-data placement result is confounded by the curriculum gating: the transition-bracketing observations are both physically informative and introduced relatively early during training. The present results therefore cannot establish whether their advantage arises from their physical location, their earlier activation, or both.

Future work should examine whether the observed placement effect persists across other curved geometries, including the convex and concave chutes studied by Greve and Hutter \cite{grevehutter1993} and flows over more complex basal topographies \cite{gray1999,pudasaini2003}. Establishing such a relationship could help develop practical guidelines for placing measurements in field applications where observations are limited and must be selected in advance. The physical and procedural effects of observation placement should also be separated by comparing transition and non-transition observations activated at the same curriculum stage. Finally, the present formulation will be extended to two dimensions, combining the curved-chute dynamics developed here with the cross-slope spreading considered in the companion study. This would provide a unified framework for PINN-based modelling of granular avalanches over curved, two-dimensional terrain.

\section*{Data availability}
Code is available on request.

\FloatBarrier

\bibliographystyle{unsrt}
\bibliography{ref}

\end{document}